\documentclass[aps,pra,nofootinbib,twocolumn]{revtex4-2}
\usepackage[colorlinks]{hyperref}
\usepackage[utf8]{inputenc}
\usepackage{graphicx}
\usepackage{multirow}
\usepackage{amssymb}
\usepackage{amsmath}
\usepackage{gensymb}

\begin{document}

\title{Searching for Scalar Field Dark Matter using Cavity Resonators and Capacitors}

\author{V.~V.~Flambaum$^{1}$}\email{v.flambaum@unsw.edu.au}
\author{B.~T.~McAllister$^{2,3}$}\email{ben.mcallister@uwa.edu.au}
\author{I.~B.~Samsonov$^1$}\email{igor.samsonov@unsw.edu.au}
\author{M.~E.~Tobar$^2$}\email{michael.tobar@uwa.edu.au}
\affiliation{$^1$School of Physics, University of New South Wales, Sydney 2052, Australia}
\affiliation{$^2$ARC Centre of Excellence For Engineered Quantum Systems\\
and ARC Centre of Excellence For Dark Matter Particle Physics, QDM Laboratory,\\
Department of Physics, University of Western Australia,
35 Stirling Highway, Crawley WA 6009, Australia}
\affiliation{$^3$ARC Centre of Excellence for Dark Matter Particle Physics, Centre for Astrophysics and Supercomputing, Swinburne University of Technology, John St, Hawthorn VIC 3122, Australia}

\begin{abstract}
We establish new experiments to search for dark matter based on a model of a light scalar field with a dilaton-like coupling to the electromagnetic field, which is strongly motivated by superstring
theory. We estimate the power of the photon signal in the process of a non-resonant scalar-photon transition and in a cavity resonator permeated by electric and magnetic fields. We show that existing cavity resonators employed in the experiments like ADMX have a low but non-vanishing sensitivity to the scalar-photon coupling. As a result, by re-purposing the results of the ADMX experiment, we find new limits on the scalar-photon coupling in the range of the scalar field masses from 2.7 to 4.2 $\mu$eV. We discuss possible modifications of this experiment, which enhance the sensitivity to the scalar field dark matter. We also propose a broadband experiment for scalar field dark matter searches based on a high-voltage capacitor. The estimated sensitivity of this experiment exceeds by nearly two orders in magnitude the sensitivity of the experiment based on molecular spectroscopy.
\end{abstract}

\maketitle

\section{Introduction}
Scalar and pseudoscalar fields are considered as leading candidates for dark matter (DM) particles. The widely known example of pseudoscalar particle is the QCD axion predicted in Refs.~\cite{PQ,Weinberg,Wilczek,Kim,SVZ,Zh1980,DFS} as an extension of the Standard Model of elementary particles. The axion-photon interaction $g_{a\gamma\gamma} a \vec E\cdot \vec B$ suggests that the axion field $a$ may be converted into photons in electric $\vec E$ or magnetic $\vec B$ field. The technique for the axion detection that utilizes this effect was proposed by P.~Sikivie in the seminal papers \cite{Sikivie83,Sikivie85}. Although the axion field has not been detected yet, the allowed values of the axion-photon coupling constant $g_{a\gamma\gamma}$ are highly constrained from a number of experiments such as ADMX \cite{ADMX1,ADMX2,ADMX3,ADMX4,ADMX5,ADMX6,ADMX7,ADMX8,ADMX9}, ORGAN \cite{ORGAN1,ORGAN2}, CAST \cite{CAST1,CAST2,CAST3}, and Light shining through wall \cite{ALPS}.

The scalar field counterpart of the axion is the dilaton $\phi$ \cite{pseudo-dilaton} that couples to the electromagnetic field through the operator $g_{\phi\gamma\gamma}\phi(B^2 - E^2)$ with the dimensionful coupling constant $g_{\phi\gamma\gamma}$. The dilaton field extension of the Standard Model is strongly motivated by the superstring theory, see, e.g., Refs.~\cite{Veneziano,DAMOUR1994,Damour1994a,DamourVeneziano,DamourVeneziano2,Kaplan_2000}. Moreover, the scalar-photon interaction serves as a window for possible physical manifestations in the chameleon models of gravity, see, e.g, Ref.~\cite{Chameleon} for a review. In contrast with the axion, the dilaton mass is much less constrained theoretically, and it represents a challenge for experimentalists to look for such particles in a wide range of values of the mass and the coupling constant. 

Both axion and dilaton fields are supposed to be light enough, so that they should exhibit wave-like properties. In particular, assuming that the scalar field makes up the main fraction of the cold non-relativistic dark matter, one concludes that such dark matter is described by a coherently oscillating background field $\phi=\phi_0 \cos \omega t$, with the frequency $\omega$ equal to the mass of this field. In Refs.~\cite{Arvanitaki,Tilburg,Flambaum1,Flambaum2,Flambaum3,Hees,Derevianko} it was shown that a coupling of this field to photons yields potentially observable effects of oscillations of fundamental constants that might be detected with atomic clocks. The scalar field dark matter was searched for with the use of optical cavities \cite{delay} and gravitational wave detectors \cite{AURIGA,Stadnik-grav,GEO600}, and the corresponding astrophysical effects were discussed in Refs.~\cite{Raffelt,Flambaum1}. Scalar-photon oscillations were studied in Ref.~\cite{oscillation}. Constraints on the scalar field dark matter coupling were found from DAMA experiment in Refs.~\cite{ScalarDAMA,Bao} and from atomic and molecular spectroscopy experiments in Refs.~\cite{Antypas2,Antypas,Aharony,Oswald2021,Tretiak}. Limits on the scalar-photon interaction within the chameleon model of gravity were found in Refs.~\cite{ChameleonADMX,ChameleonOthers}. Recently, it was proposed to search for the scalar dark matter with photonic, atomic, and mechanical oscillators \cite{Oscillators}.

In this paper, we consider the possibility of detection of the scalar field dark matter using cavity resonator haloscope techniques analogous to the axion dark matter model \cite{Sikivie83,Sikivie85}. We show that the axion-detecting cavity experiments like ADMX have a low but non-vanishing sensitivity to the scalar field coupling $g_{\phi\gamma\gamma}$. This allows us to find new limits on this constant by simply re-purposing the results of the ADMX experiment \cite{ADMX6,ADMX9}. Further improvements of these limits would require some modifications of the existing cavities employed in such experiments. We propose some modifications of this experiment with various configurations of electric and magnetic fields to maximize the sensitivity.

Although cavity resonators may have a good sensitivity to the scalar-photon coupling $g_{\phi\gamma\gamma}$, they usually allow one to explore a relatively narrow band of frequencies while the range of allowed masses of DM particles is poorly limited. In this respect, broadband experiments are necessary for studying the scalar field dark matter model. We propose one such experiment that utilizes the effect of transformation of the scalar field into a photon in a strong electric field inside a capacitor. We show that light scalar field dark matter should induce detectable oscillating electric field in the capacitor. We expect that this technique should improve existing limits on $g_{\phi\gamma\gamma}$ found in Refs.~\cite{Antypas2,Antypas,Aharony,Oswald2021,Tretiak} from atomic and molecular spectroscopy experiments. 
 
The rest of the paper is organized as follows. In Sec.~\ref{SecTheory}, we develop analytical expressions for the power of the photon signal from the scalar-photon transformation. Here we consider both non-resonant transition in electric and magnetic fields and transformation inside cavity resonators. We provide also numerical estimates for the power of this transformation in a cavity similar to the one used in the ADMX experiment. In Sec.~\ref{SecExperiments}, we find new constraints on the scalar-photon coupling constant by re-purposing the results of the ADMX experiment and propose new experiments with enhanced sensitivity to the scalar field dark matter. Here we present also a new experimental approach for the broadband detection of the scalar field dark matter with the use of a capacitor sourced by a high voltage DC power unit and compare the projected sensitivity with the limits from other experiments. Section \ref{SecExperiments} is devoted to a summary and discussion of the results of this work.

In this paper, we use natural units with $\hbar = c = \epsilon_0 = \mu_0 =1$. The results of numerical estimates will be represented in SI units as well.


\section{Scalar-photon transformation}
\label{SecTheory}

\subsection{Scalar-photon interaction}

In this section, we introduce the dilaton-like interaction of the scalar field with the electromagnetic field and derive the corresponding modifications of the Maxwell equations. We will show that the scalar field dark matter interaction with the permanent electric and magnetic field serves as a source for electromagnetic waves.

\subsubsection{Interaction with background scalar field}

In vacuum, the interaction of the electromagnetic field $F_{\mu\nu}$ with a real scalar field $\phi$ is described by the Lagrangian
\begin{equation}
\label{L1}
    {\cal L} =-\frac14 F_{\mu\nu}F^{\mu\nu} -
    \frac{g_{\phi\gamma\gamma}}4 \phi F_{\mu\nu}F^{\mu\nu}
    \,,
\end{equation}    
with a dimensionful coupling constant $g_{\phi\gamma\gamma}$.
The scalar field $\phi$ here is considered as a background field obeying the Klein-Gordon equation $(\Box+m_\phi^2)\phi = 0$ with mass $m_\phi$. In this paper, we will consider a plain-wave solution for the scalar field
\begin{equation}
\label{eq45}
    \phi ={\rm Re}[\phi_0 e^{i(\vec p \cdot \vec x - \omega t)}]
\end{equation}
with momentum $\vec p$ and energy $\omega = \sqrt{p^2 + m_\phi^2}$. 

In general, dark matter particles may be produced either in decays of standard model particles or may have a primordial origin. In the former case these particles are usually ultrarelativistic while in the latter case they contribute to the cold dark matter. If these particles saturate the local dark matter density $\rho_{\rm DM}=0.45\,{\rm GeV/cm}^3$, the amplitude of the scalar field (\ref{eq45}) may be written as 
\begin{equation}
    \phi_0 = \frac{\sqrt{2\rho_{\rm DM}}}{m_\phi}\,.
    \label{phi0}
\end{equation}
The momentum of such particles may be written as
$ \vec p = \omega \vec\beta$, with $\beta = 10^{-3}c$ being the virial velocity in our Galaxy.

In this section, however, we will consider the general case with arbitrary DM particle velocity. The ultrarelativistic case will be discussed at the end of this section with a reference to the CAST experiment \cite{CAST1,CAST2,CAST3} which may detect scalar DM particles emitted from the Sun.

\subsubsection{Modifications of Maxwell's equations in a medium}

In a medium with the dielectric constant $\epsilon$ and magnetic susceptibility $\mu$, the Lagrangian (\ref{L1}) becomes
\begin{equation}
    {\cal L}=
    \frac12(\epsilon \vec E^2 - \frac1\mu \vec B^2)
    +\frac12g_{\phi\gamma\gamma} \phi( \vec E^2 -  \vec B^2)
    \,,
\end{equation}
where $\vec E = -\nabla \Phi - \partial_t \vec A$ and $\vec B = \nabla\times \vec A$ are electric and magnetic fields, respectively. The corresponding modifications of Maxwell's equations are
\begin{subequations}
\begin{eqnarray}
\nabla \cdot (\epsilon \vec E + g_{\phi\gamma\gamma} \phi \vec E) &=& 0\,,\label{eq10a}\\
\nabla \times (\mu^{-1} \vec B + g_{\phi\gamma\gamma} \phi \vec B ) - \partial_t (\epsilon \vec E + g_{\phi\gamma\gamma} \phi \vec E) &=& 0\,,\label{eq10}\\
\nabla \times \vec E + \partial_t \vec B &=& 0\,,\\
\nabla \cdot \vec B &=& 0\,.
\end{eqnarray}
\end{subequations}

Consider background static electric $\vec E_0$ and magnetic $\vec B_0$ fields. Equations (\ref{eq10a}) and (\ref{eq10}) suggest that the scalar field $\phi$ interacting with these background fields serves as a source of effective electric charge and current densities:
\begin{subequations}
\label{current}
\begin{align}
    \rho_\phi &= -g_{\phi\gamma\gamma}\nabla \cdot(\phi \vec E_0)\,,\\
    \vec j_\phi &=  g_{\phi\gamma\gamma} \vec E_0 \partial_t \phi
    -g_{\phi\gamma\gamma} \nabla\times (\phi  \vec B_0 ) \,.
\end{align}
\end{subequations}
With these sources, Eqs.~(\ref{eq10a}) and (\ref{eq10}) may be cast in the form
\begin{subequations}
\label{Maxw}
\begin{eqnarray}
\nabla\cdot(\epsilon \vec E) &=& \rho_\phi\,,\\
\nabla \times (\mu^{-1} \vec B) - \partial_t(\epsilon \vec E) &=&\vec j_\phi\,.
\label{7b}
\end{eqnarray}
\end{subequations}

Note that $\rho_\phi$ and $j_\phi$ satisfy the continuity equation, $\partial_t \rho_\phi + \nabla\cdot \vec j_\phi = 0$.

\subsection{Non-resonant transformation power in static electric and magnetic fields}

In this section, we will find a solution of Eqs.~(\ref{Maxw}) that will be applied for calculation of the scalar-photon transformation power. This derivation is similar to the one for the axion-photon transformation presented in Ref.~\cite{Sikivie2020}.

In what follows, we assume that $\epsilon$ and $\mu$ are constant over the space and time. Then, in the Lorentz gauge, $\epsilon\mu \partial_t \Phi + \nabla\cdot \vec A =0$, the Maxwell equations (\ref{Maxw}) imply
\begin{eqnarray}
\label{EqPhi}
    (-\nabla^2 + \epsilon\mu \partial_t^2 )\Phi &=& \epsilon^{-1}\rho_\phi \,,\\
    (-\nabla^2 + \epsilon\mu \partial_t^2 )\vec A &=& \mu \vec j_\phi \,,
    \label{eq47}
\end{eqnarray}
with $\rho_\phi$ and $j_\phi$ given by Eqs.~(\ref{current}).

For deriving the power of radiation from scalar-to-photon transformation it is sufficient to consider only the equation for the vector potential (\ref{eq47}) because it is possible to show that the contribution with $\rho_\phi$ drops out from the Pointing vector at large distance from the source (in the wave zone), see, e.g., \cite{LL2}. 

We will look for a solution of Eq.~(\ref{eq47}) within the ansatz $\vec A(x,t) = {\rm Re}[\vec A(x)e^{-i\omega t}]$ with $\vec A(x)$ obeying
\begin{subequations}
\begin{align}
    &(-\nabla^2 - \epsilon\mu \omega^2)\vec A(x) = \mu \vec j_\phi(x)\,,\label{eq15a}\\
    &\vec j_\phi(x) = -g_{\phi\gamma\gamma}
\phi_0 e^{i\vec p\cdot\vec x}     
    [i\omega \vec E_0 + (\nabla+ i\vec p)\times \vec B_0 ]\,.
    \label{eq15}
\end{align}
\end{subequations}
Here the plane wave solution for the scalar field (\ref{eq45}) has been employed. Making use of the Fourier transform for the external magnetic field, 
$\vec B_0(x)=\int d^3x \, e^{i\vec q \cdot\vec x} \vec B_0(q)$, the last term in Eq.~(\ref{eq15}) may be written as
$(\nabla + i\vec p ) \times \vec B_0(x) = i\int d^3x \, e^{i\vec q \cdot\vec x}(\vec q + \vec p)\times \vec B_0(q)$. Here $\vec p$ represents the scalar field momentum while $\vec q$ is the momentum carried by the external magnetic field $\vec B_0$. We will keep both these momenta because either of these terms may be leading. In particular, in helioscope experiments like CAST, the scalar field momentum is supposed to be in the keV range, and the condition $p\gg q$ is satisfied. In the search of ultralight ($m\ll 10^{-6}$ eV) Galactic halo scalar dark matter, the opposite condition may be satisfied, $p\ll q$. In microwave cavities like the one in the ADMX experiment, these two terms may be comparable, $p\sim q$.

The solution of Eq.~(\ref{eq15a}) may be written in terms of the Green's function,
\begin{equation}
    \vec A(x) = \frac\mu{4\pi} \int_V d^3x' \frac{e^{i k|\vec x-\vec x'|}}{|\vec x - \vec x'|} \vec j_\phi(x')\,,
\end{equation}
where $ k = \sqrt{ \epsilon\mu} \omega$. The integration is performed over the volume $V$ with non-vanishing electric $\vec E_0$ and magnetic $\vec B_0$ fields. At large distance, $r\equiv  |\vec x|\gg V^{1/3}$, this solution is approximated as
\begin{equation}
    \vec A(\vec x) = \mu \frac{e^{i k r}}{4\pi r}\vec j_\phi(k)+ O \left(\frac1{r^2}\right)\,,
\end{equation}
where 
\begin{align}
    \vec j_\phi(k) &= \int_V d^3x\, e^{-i\vec k \cdot \vec x} \vec j_\phi(x)\\
    &=-ig_{\phi\gamma\gamma} \phi_0 \int_V d^3x \, e^{i(\vec p - \vec k)\vec x}[(\vec p - i\nabla)\times\vec B_0 + \omega\vec E_0]\,.
    \nonumber
\end{align}
Here $\vec p = \omega\vec\beta$, $\vec x = r \vec n$, and $\vec k = k \vec n $, with a unit vector $\vec n$. 

The power of electromagnetic radiation per solid angle in a direction $\vec n$ is \cite{Sikivie2020}
\begin{equation}
    \frac{dP}{d\Omega} = \lim_{r\to\infty}
    \langle \vec n \cdot(\vec E\times \vec H)\rangle r^2 = \frac{\mu k \omega}{32\pi^2} |\vec n\times \vec j_\phi(k)|^2\,,
    \label{eq22}
\end{equation}
where $\langle\ldots\rangle$ denotes the time average. The power of the scalar-photon transition is obtained upon integration of Eq.~(\ref{eq22}) over the solid angle $\Omega$,
\begin{align}
\label{Power}
    &P =  \frac{g_{\phi\gamma\gamma}^2 \rho_{\rm DM}}{16\pi^2\epsilon}\int d^3k\, \delta(|\vec k|-\omega)
    \\&\times\left| \int_V d^3x \, e^{i(\vec p - \vec k)\cdot \vec x }
\vec n \times [\vec E_0 + (\vec\beta -i \omega^{-1}\nabla)\times \vec B_0 ] \right|^2.
\nonumber
\end{align}
Here we made use of the relation (\ref{phi0}). Using a similar relation for the energy flux of the scalar field, $\vec{\cal P} = \frac12 \phi_0^2\omega^2 \vec\beta$, we find also the differential cross section for the scalar field to photon transformation
\begin{align}\label{CrossSection}
&\frac{d\sigma}{d\Omega} = \frac1{|\vec{\cal P}_\phi|}\frac{dP}{d\Omega}= g_{\phi\gamma\gamma}^2\frac{\mu k \omega}{16\pi^2\beta} \\&
\times\left|
\int_V d^3x\, e^{i(\vec p - \vec k)\cdot \vec x }
\vec n \times[\vec E_0 +(\vec\beta - i\omega^{-1}\nabla)\times \vec B_0 ]
\right|^2.\nonumber
\end{align}

The results for the differential cross section (\ref{CrossSection}) and power (\ref{Power}) are very similar to the corresponding expressions for the axion-photon transformation found in Refs.~\cite{Sikivie83,Sikivie85}. However, the roles of external electric and magnetic fields are swapped now. In particular, when the electric field is vanishing, that is the case in many cavity haloscope experiments, the cross section (\ref{CrossSection}) depends on the angle between applied magnetic field $\vec B_0$ and the scalar field velocity, $|\vec B_0\times \vec \beta|^2=B^2_0 \beta^2 \sin^2\theta$. As a result, the factor $\sin^2\theta$ should result in daily and annual modulation of the expected signal from the scalar field to photon transformation. The factor $\beta^2\sim 10^{-6}$ yields the suppression of this effect, as compared with the axion-photon transformation cross section \cite{Sikivie83,Sikivie85}. 

\begin{figure}
    \centering
    \includegraphics[width=8.5cm]{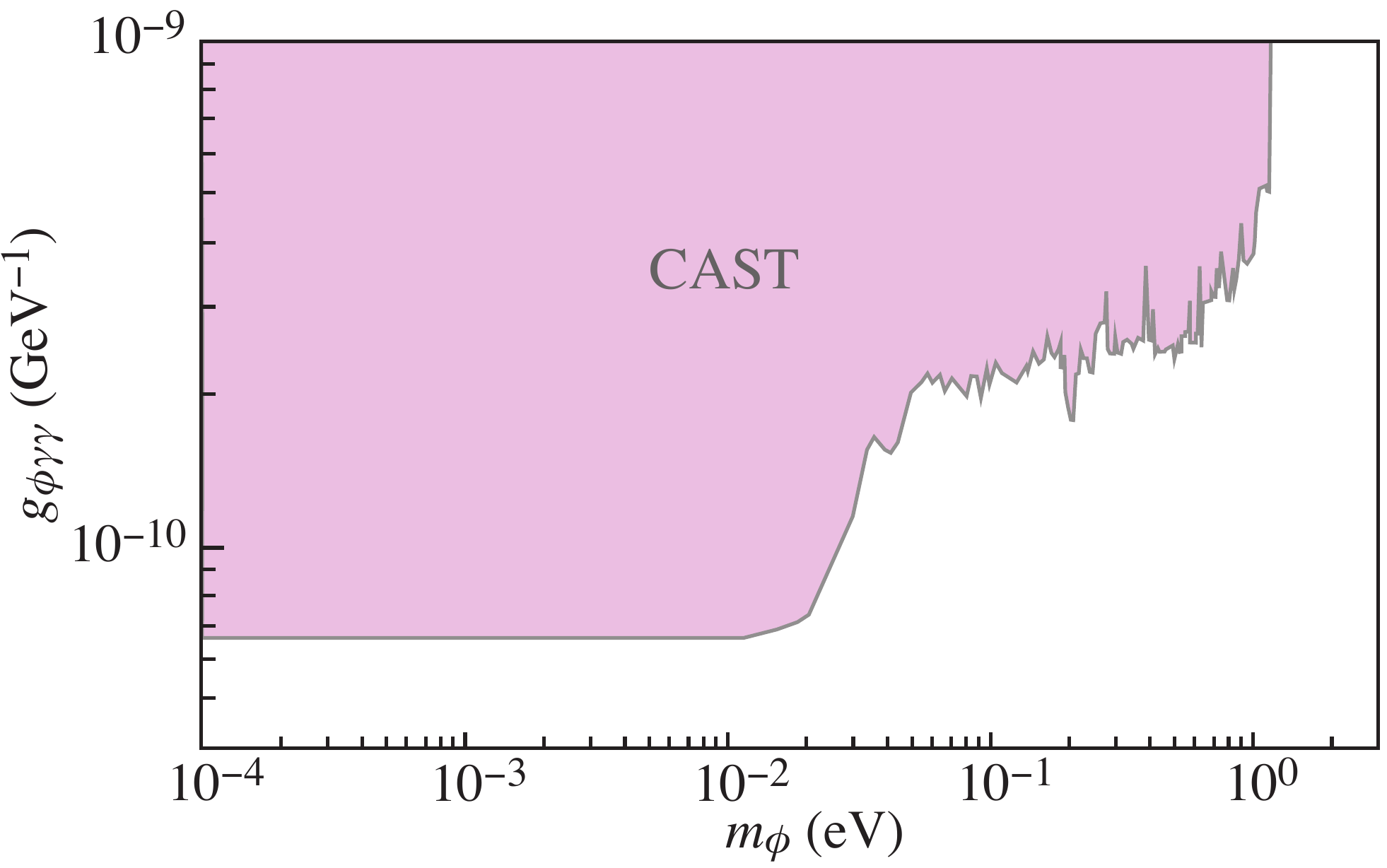}
    \caption{Upper limits on the coupling $g_{\phi\gamma\gamma}$ derived with a graphical accuracy from the results of the CAST experiment \cite{CAST3}.}
    \label{FigCAST}
\end{figure}

Helioscopes, however, can detect scalar and pseudoscalar particles with equal efficiency. Indeed, the CAST experiment \cite{CAST1,CAST2,CAST3} aims at detecting axions produced by the Sun with energies of order 1 keV. In contrast with the galactic halo dark matter, such axions are ultrarelativistic with $\beta\approx1$. Moreover, since this helioscope is based on a moving platform and aims at the Sun, the applied magnetic field is always perpendicular to the axion or scalar field momentum, $\sin\theta\approx 1$. Thus, all constraints found by the CAST experiment \cite{CAST3} on the axion-photon coupling constant $g_{a\gamma\gamma}$ literally apply to the scalar field coupling $g_{\phi\gamma\gamma}$. The corresponding limits on $g_{\phi\gamma\gamma}$ are given in Fig.~\ref{FigCAST}.

Note that in the recent paper \cite{Sokolov} it was shown that the interaction similar to the one in Eqs.~(\ref{current}) and (\ref{Maxw}) appears also in the axion electrodynamics with magnetic monopoles. Therefore, non-detection of solar axions in the CAST experiment is consistent with non-observation of magnetic monopoles.

The opposite case with vanishing external magnetic field and non-zero electric field corresponds to the scalar field to photon transformation in matter. In this case, the role of the electric field $E_0$ is played by the internal Coulomb fields in atoms and molecules. Possibility of detection of this effect in the DAMA experiment was discussed in Refs.~\cite{ScalarDAMA,Bao}. In contrast with the axion case, however, we expect no daily and annual modulation of the signal because the power of produced electromagnetic field is independent of $\beta\sin\theta$, see Eq.~(\ref{Power}).

Note also that astrophysical constraints on the scalar field coupling $g_{\phi\gamma\gamma}$ were studied in a recent paper \cite{Raffelt}.

\subsection{Transformation in a cavity resonator}

In Refs.~\cite{Sikivie83,Sikivie85} it was shown that axion dark matter could be effectively detected with the use of cavity resonators. Such cavity resonators have been employed, e.g., in ADMX \cite{ADMX1,ADMX2,ADMX3,ADMX4,ADMX5,ADMX6,ADMX7,ADMX8,ADMX9} and ORGAN \cite{ORGAN1,ORGAN2} experiments. In this section, we will show that the cavity resonator technique appears efficient for measuring the scalar-photon coupling $g_{\phi\gamma\gamma}$ as well. We will calculate the power of the photon signal from the scalar-photon transformation in permanent electric and magnetic fields in such cavities.

\subsubsection{Estimates of the photon signal power}
 
Consider a cavity of volume $V$ with static electric $\vec E_0$ and magnetic $\vec B_0$ fields. The vector potential in the cavity is expanded over the cavity eigenmodes $\vec e_\alpha$, ($\Phi=0$ gauge is assumed \footnote{In general, this gauge is inconsistent with Eq.~(\ref{EqPhi}), since a non-vanishing field $\Phi$ is produced by the effective charge density $\rho_\phi$. For cold dark matter described by the scalar field (\ref{eq45}) the effective charge density may however be ignored, $\rho_\phi=0$, as compared with the effective current $\vec j_\phi$, and the gauge $\Phi=0$ may be imposed. Indeed, the effective charge density appears in the following corollary of Maxwell's equations (\ref{Maxw}): $\nabla^2 \vec E -\epsilon\mu \partial_t^2 \vec E = \mu \partial_t \vec j_\phi + \epsilon^{-1}\nabla\rho_\phi$. For a slowly varying electric field $\vec E_0$ and dielectric constant $\epsilon$, the last term here is suppressed by the factor $\beta^2=10^{-6}$ with $\vec \beta = \vec p/\omega$ the virial velocity of DM particles.})
\begin{equation}
    \vec A(x,t) = \sum_\alpha \vec e_\alpha(x)\psi_\alpha(t)\,,
    \label{Ae}
\end{equation}
which satisfy
\begin{subequations}
\label{eq62}
\begin{eqnarray}
\nabla\cdot(\epsilon \vec e_\alpha) &=& 0\,,\\
\nabla\times (\mu^{-1}\nabla\times \vec e_\alpha) - \epsilon\omega_\alpha^2 \vec e_\alpha &=& 0\,,\label{eq60}\\
\vec n \times \vec e_\alpha |_S &=& 0\,,\label{61}\\
\int_V d^3x\, \epsilon\, \vec e_\alpha(x) \vec e_\beta(x) &=& \delta_{\alpha\beta}\,,
\end{eqnarray}
\end{subequations}
where $S$ is the boundary of the cavity and $\vec n$ is the unit vector normal to its surface, $\omega_\alpha$ are the eigenfrequencies. Substituting the vector potential (\ref{Ae}) into Eq.~(\ref{7b}) and making use of Eqs.~(\ref{eq62}) we find that the amplitude $\psi_\alpha(t)$ obeys the damped driven harmonic oscillator equation
\begin{equation}
        \left(\frac{d^2}{dt^2} + \gamma_\alpha \frac{d}{dt} + \omega_\alpha^2\right) \psi_\alpha(t) = -g_{\phi\gamma\gamma} \phi_0 {\rm Re}[ e^{-i\omega t}F(p)] 
        \label{eq27}
\end{equation}
with
\begin{align}
\label{FP}
        F(p) = \int_V d^3x\, e^{i\vec p\cdot \vec x} \vec e_\alpha \cdot[i\omega \vec E_0 +(\nabla + i\vec p)\times \vec B_0]\,.
\end{align}
Here $\gamma_\alpha$ is the damping constant related to the cavity quality factor as $Q_\alpha = \omega_\alpha/\gamma_\alpha$. The term with the dissipation in Eq.~(\ref{eq27}) is added manually \footnote{It is possible to show that the same result may be found in a more systematic way by adding to the Maxwell equations (\ref{Maxw}) the terms with external charges and currents arising in the cavity walls. Similar derivation based on the complex Poynting theorem was reported in Ref.~\cite{TMG} for the axion electrodynamics.}.

In resonance, $\omega_\alpha=\omega\approx m_\phi$, the steady state solution of Eq.~(\ref{eq27}) reads
\begin{equation}
    \psi_\alpha(t) = - \frac{g_{\phi\gamma\gamma} \phi_0}{ \omega \gamma_\alpha}{\rm Re}[ i e^{-i\omega t} F(p)]\,,
    \label{psi}
\end{equation}
and, making use of the integration by parts with the boundary condition (\ref{61}), Eq.~(\ref{FP}) may be cast in the form
\begin{equation}
        F(p)=\frac1{\psi_\alpha}\int_V d^3x\,
        e^{i\vec p\cdot \vec x}(\vec B_0 \cdot \vec B_\alpha + \vec E_0 \cdot \vec E_\alpha)
        \,.
\end{equation}
Here $\vec E_\alpha = -\psi'_\alpha \vec e_\alpha$ and $\vec B_\alpha = \psi_\alpha \nabla \times \vec  e_\alpha$ are the electric and magnetic fields of the $\alpha$ eigenmode in the cavity, respectively. In terms of the function (\ref{FP}) these fields read
\begin{equation}
\begin{aligned}
\vec E_\alpha &= g_{\phi\gamma\gamma} \phi_0  \gamma_\alpha^{-1} {\rm Re}\left[ e^{-i\omega t} 
        F(p)\right] \vec e_\alpha \,,\\
\vec B_\alpha &= - \frac{g_{\phi\gamma\gamma} \phi_0}{ \omega\gamma_\alpha} {\rm Re}\left[i e^{-i\omega t}  F(p)\right]\nabla\times \vec e_\alpha\,.
\end{aligned}
\end{equation}
Making use of these solutions we find the time-averaged power from conversion of the scalar field into the $\alpha$ mode of the cavity
\begin{align}
P &= \frac{\gamma_\alpha}2 \int_V d^3x\left\langle \epsilon \vec E_\alpha \cdot \vec E_\alpha
+\frac1{\mu} \vec B_\alpha\cdot \vec B_\alpha
\right\rangle \nonumber\\
&=\frac{g_{\phi\gamma\gamma}^2 \phi_0^2}{2\gamma_\alpha} 
|F(p)|^2\,.
\label{P32}
\end{align}
Remembering that $Q_\alpha = \omega/\gamma_\alpha$ is the cavity quality factor in the mode $\alpha$ and $\phi_0^2 = 2\rho_{\rm DM}/m_\phi^2 $, we represent Eq.~(\ref{P32}) in the conventional form
\begin{equation}
\label{P33}
    P = \frac1{m_\phi}g_{\phi\gamma\gamma}^2 \rho_{\rm DM}( B_0^2 + E_0^2) V C_\alpha Q_\alpha\,,
\end{equation}
where
\begin{equation}
C_\alpha 
     =\frac1{(B_0^2 + E_0^2) V}\frac{\left|\int_V d^3x\, e^{i\vec p\cdot \vec x} (\vec B_0 \cdot \vec B_\alpha + \vec E_0 \cdot \vec E_\alpha) \right|^2}{\frac12 \int_V d^3x(\mu^{-1}\vec B_\alpha\cdot\vec B_\alpha + \epsilon \vec E_\alpha \cdot \vec E_\alpha)}
\label{formfactor}
\end{equation}
is the form factor which quantifies the coupling strength of cavity eigenmode $\alpha$ to the external electric $\vec E_0$ and magnetic $\vec B_0$ fields.

\subsubsection{Coupling to external magnetic field}
In axion detecting cavities like ADMX or ORGAN the strong magnetic fields are applied while the electric field is vanishing, $\vec E_0=0$. In this case, the general expressions (\ref{P33}) and (\ref{formfactor}) become
\begin{align}
    P&=\frac1{m_\phi} g_{\phi\gamma\gamma}^2 \rho_{\rm DM} B_0^2  V C_\alpha Q_\alpha \,,\label{P37}\\
    C_\alpha 
     &=\frac1{B_0^2  V}\frac{\left|\int_V d^3x\, e^{i\vec p\cdot \vec x} \vec B_0 \cdot \vec B_\alpha  \right|^2}{ \int_V d^3x\, \mu^{-1}\vec B_\alpha\cdot\vec B_\alpha }\,.
\label{CB}
\end{align}
Note that the factor $e^{i\vec p\cdot \vec x}$ may be omitted if the cavity is much smaller than the scalar field de Broglie wavelength. 

The radiation power (\ref{P37}) may be represented in physical units (watts):
\begin{eqnarray}
    P &=& 1.3\times 10^{8}{\rm W} \, 
    \left(\frac{g_{\phi\gamma\gamma}}{{\rm GeV}^{-1}} \right)^2
    \left(\frac{3\mu{\rm eV}}{m_\phi} \right)
    \left(\frac{\rho_{\rm DM}}{0.45 {\rm GeV/cm}^3} \right)
    \nonumber\\&&\times
    \left(\frac{B_0}{7.6\rm T} \right)^2 
    \left(\frac{V}{136\rm L} \right)
    \left(\frac{C_\alpha}{0.4} \right)
    \left(\frac{Q_\alpha}{30000} \right).
\label{PowerADMX}
\end{eqnarray}
This expression is very similar to the power of axion-to-photon conversion estimated for the ADMX experiment \cite{ADMX6}. These expressions coincide upon the substitution $g_{\phi\gamma\gamma}\to g_{a\gamma\gamma}$ and $m_\phi \to m_a$. However, as will be shown in Sec.~\ref{SecADMX}, the value of the form factor (\ref{CB}) appears much smaller than that for the axion.

\subsubsection{Coupling to external electric field}

When the external magnetic field is vanishing, $\vec B_0=0$, the general expressions (\ref{P33}) and (\ref{formfactor}) turn into 
\begin{align}
    P&=\frac1{m_\phi} g_{\phi\gamma\gamma}^2 \rho_{\rm DM} E_0^2  V C_\alpha Q_\alpha \,,\label{P38}\\
    C_\alpha 
     &=\frac1{E_0^2  V}\frac{\left|\int_V d^3x\, e^{i\vec p\cdot \vec x} \vec E_0 \cdot \vec E_\alpha  \right|^2}{ \int_V d^3x\, \epsilon \vec E_\alpha\cdot\vec E_\alpha }\,.
\label{P38-formfactor}
\end{align}
The latter equation shows that if the homogeneous external electric field $\vec E_0$ is directed along the $z$ axis of a cylindrical cavity, the dominant cavity eigenmode is TM$_{010}$. The corresponding form factor reads $C_\alpha = 0.69$ for a perfect cylindrical cavity (see, e.g., \cite{Sikivie2020}), or $C_\alpha=0.4$ for the cavity employed in the ADMX experiment \cite{ADMX6}.
Assuming the latter value for the form factor, we represent the power (\ref{P38}) in physical units:
\begin{equation}
\begin{aligned}
    P &= 38{\rm W}\left(\frac{g_{\phi\gamma\gamma}}{{\rm GeV}^{-1}} \right)^2
    \left(\frac{3\mu{\rm eV}}{m_\phi} \right)
    \left(\frac{\rho_{\rm DM}}{0.45 {\rm GeV/cm}^3} \right)
    \\&\times
    \left(\frac{E_0}{1\, 
    \rm MV/m} \right)^2 
    \left(\frac{V}{136\,\rm L} \right)
    \left(\frac{C_\alpha}{0.4} \right)
    \left(\frac{Q_\alpha}{30000} \right).
\end{aligned}
\label{PowerElectric}
\end{equation}
Note that here we assumed reference values for the cavity volume and quality factor as in the ADMX experiment \cite{ADMX6,ADMX9}.

Comparing Eq.~(\ref{PowerElectric}) with Eq.~(\ref{PowerADMX}) we conclude that the cavity resonators with magnetic field are in general much more efficient for detection of scalar field dark matter.

\subsubsection{Signal-to-noise ratio}

The detection technique of the scalar field dark matter with cavity resonator is very similar to the corresponding axion detection experiments like ADMX \cite{ADMX8}. Therefore, the signal-to-noise ratio is given by the standard expression (see, e.g., \cite{Sikivie2020}) 
\begin{equation}
    {\rm SNR} = \frac{P}{k_B T}\sqrt{\frac{t}{b}}\,,
\end{equation}
where $P$ is the expected signal power (\ref{P33}), $T$ is the total noise temperature, $t$ is the integration time and $b\equiv  f/Q$ is the cavity bandwidth. 


\section{Experiments proposals and constraints}
\label{SecExperiments}

In this section, we analyse the sensitivity of axion detection experiments to the scalar-photon coupling $g_{\phi\gamma\gamma}$. As will be shown below, it is possible to find new constraints on this coupling just by re-purposing the results of the ADMX experiment. More over, we propose new experiments dedicated to the detection of the scalar field dark matter. We divide them into two large classes corresponding to resonant cavities and broadband detection techniques, respectively.

\subsection{Broadband detection with a capacitor}
\label{SecBroadband}

In this section we show that a capacitor under a strong DC electric field may serve as a broadband experiment to search for scalar field dark matter.

\begin{figure}
    \includegraphics[width=8.5cm]{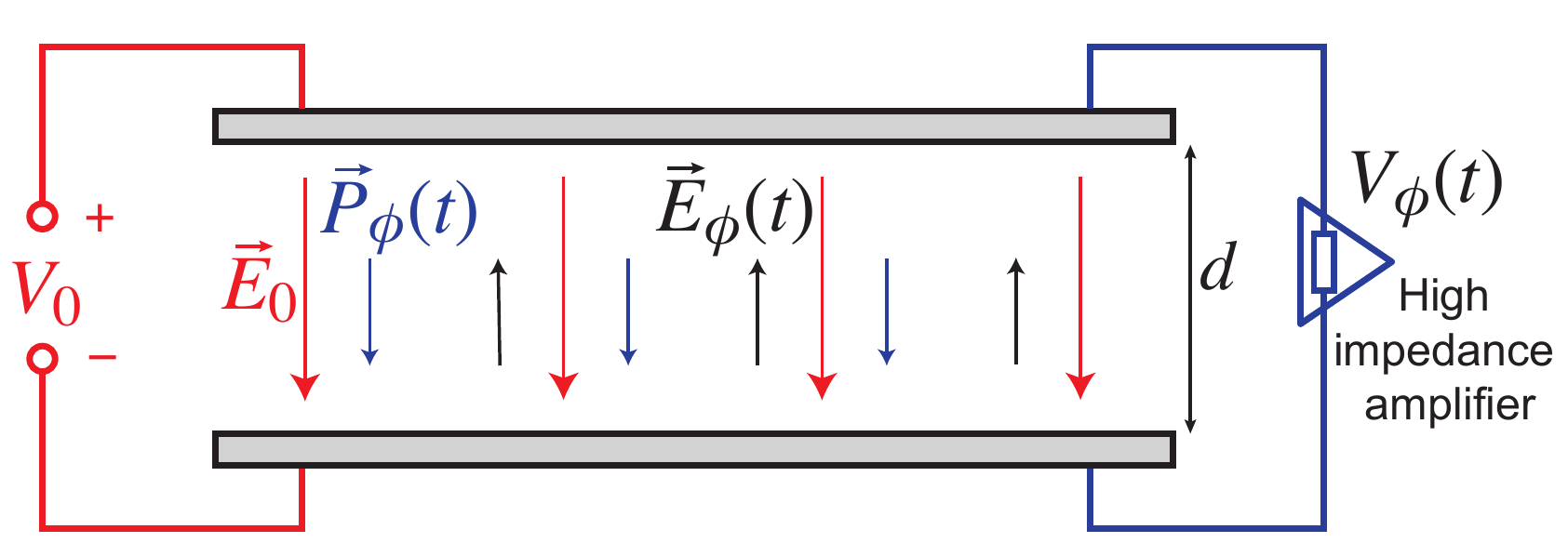}
    \caption{A sketch of a capacitive broad band experiment for scalar field dark matter search. The capacitor is charged by a high voltage $V_0$ producing a static electric field $\vec E_0$ inside the capacitor. Scalar field dark matter interacting with $\vec E_0$ creates the effective polarization $\vec P_\phi$ and oscillating electric field $\vec E_\phi$. This produces an alternating voltage $V_\phi = E_\phi d $ registered through a high impedance amplifier.}
    \label{FigCap}
\end{figure}

Consider a capacitor with a plate separation $d$ charged by an external source of a high voltage $V_0$ as in Fig.~\ref{FigCap}. The corresponding permanent electric field strength inside the capacitor is $E_0 = V_0/d$. The interaction of the scalar field dark matter (\ref{eq45}) with this electric field is described by Eqs.~(\ref{Maxw}). With no external magnetic field, $\vec B_0=0$, these equations may be cast in the form 
\begin{equation}
\label{Eq33}
\begin{aligned}
    \nabla\cdot(\epsilon \vec E + \vec P_\phi) &=0\\
    \nabla(\mu^{-1}\vec B) -\partial_t (\epsilon \vec E + \vec P_\phi) &=0\,,
\end{aligned}
\end{equation}
with 
\begin{equation}
    \vec P_\phi = g_{\phi\gamma\gamma} \phi \vec E_0
\end{equation}
being the effective polarization vector due to the scalar field interaction with the electric field $\vec E_0$. The electric field $\vec E$ may be expanded as $\vec E = \vec E_0 + \vec E_\phi$, with the background field $\vec E_0$. As follows from Eqs.~(\ref{Eq33}), the field $\vec E_\phi$ reads
\begin{equation}
    \vec E_\phi = -\epsilon^{-1}\vec P_\phi =- g_{\phi\gamma\gamma}\phi \epsilon^{-1} \vec E_0\,.
\end{equation}
This field $\vec E_\phi$ induces an AC voltage on the plates of the capacitor,
\begin{equation}
\label{Vphi}
    V_\phi = \vec E_\phi\cdot \vec d = -g_{\phi\gamma\gamma}\epsilon^{-1}  \phi V_0\,.
\end{equation}
Note that this effect depends solely on the applied voltage $V_0$ and may be observed practically in any capacitor. 

Assuming that the size of the capacitor is much smaller than the de Broglie wavelength of the scalar field, Eq.~(\ref{eq45}) may be written as
$\phi = \phi_0 \cos \omega t$. In this case, Eq.~(\ref{Vphi}) implies the following rms voltage due to the scalar field dark matter
\begin{equation}
\label{signal}
    \langle V_\phi \rangle = \frac1{\sqrt2}g_{\phi\gamma\gamma}\epsilon^{-1}V_0 \phi_0 = g_{\phi\gamma\gamma}\epsilon^{-1} V_0 \frac{\sqrt{\rho_{\rm DM}}}{m_\phi}\,,
\end{equation}
where we made use of Eq.~(\ref{phi0}) for the scalar field amplitude $\phi_0$. Equation (\ref{signal}) shows that the signal is linear in the scalar-photon coupling $g_{\phi\gamma\gamma}$. Thus, this experiment may have an advantage against other experiments where the signal is quadratic in $g_{\phi\gamma\gamma}$.

\begin{figure}
    \includegraphics[width=8.8 cm]{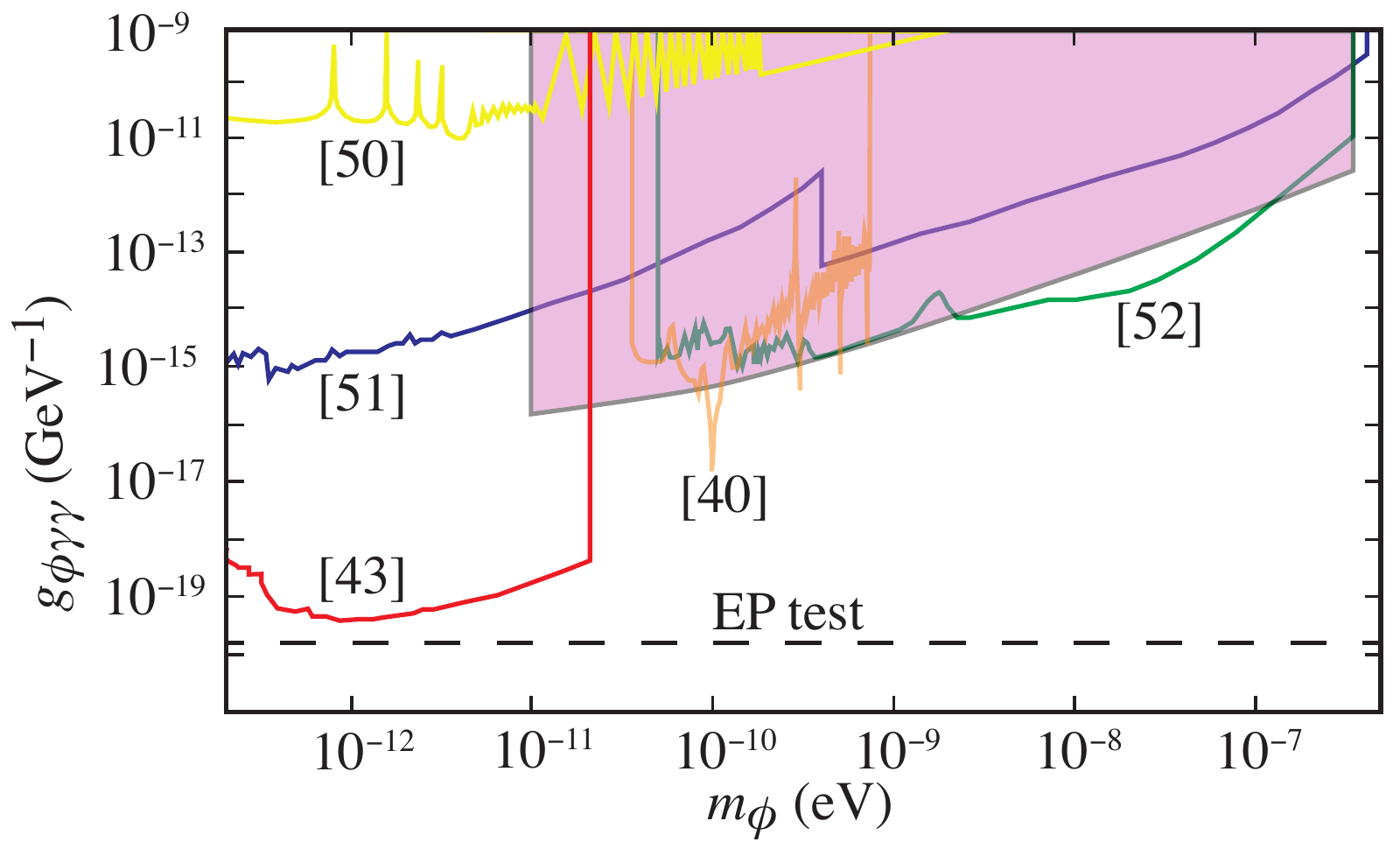}
    \caption{Projected sensitivity of the broadband detection capacitor-based experiment (pink region) and its comparison with the constraints on the scalar-photon coupling $g_{\phi\gamma\gamma}$ from other experiments \cite{delay,GEO600,Aharony,Oswald2021,Tretiak}. It is assumed that the capacitor is sourced with a 600 kV voltage and has SNR=1. The dashed line represents the constraints from the equivalence principle \cite{EP1,EP2,Stadnik-equivalence}.
    }
    \label{FigCapSens}
\end{figure}

In our numerical estimates we assume the applied voltage $V_0 = 600$ kV. This voltage may be produced, e.g., by a commercially available X-rays power supply \cite{600kV}. For the signal detection a low noise RF-amplifier may be used such as HFC 50 D/E \cite{amplifier} with spectral noise floor $\sqrt{S_V} \geq 0.2\,{\rm nV}/\sqrt{\rm Hz}$ below 50 MHz. Numerically, the spectral noise of this amplifier (in volts) may be modelled by the function
\begin{equation}
    \sqrt{S_V} = \sqrt{\frac{7.4185\times 10^{-14}}{f^{1.12}}+\frac{9.252\times 10^{-19}}{f^{0.176}}}\,,
\end{equation}
with $f$ the signal frequency measured in hertz. With the noise described by this function and the scalar field signal (\ref{signal}), we find the signal-to-noise ratio for the scalar field  dark matter detection experiment sketched in Fig.~\ref{FigCap}:
\begin{equation}
{\rm SNR}= g_{\phi\gamma\gamma} V_0\sqrt{\frac{\rho_{\rm DM}}{ S_{V}}}\left(\frac{10^6t}{f_\phi}\right)^{\frac{1}{4}}\,.
\label{SNRvolt}
\end{equation}
Here we assume that the virialized dark matter has a bandwidth, $\Delta f_\phi$, of order of a part in $10^6$, so $\Delta f_\phi\sim\frac{f_\phi}{10^6}$, and the measurement time $t$ is greater than the scalar field coherence time so that $t>\frac{10^6}{ f_{\phi}}$. For measurement times of $t<\frac{10^6}{ f_{\phi}}$ we substitute $\left(\frac{10^6t}{ f_{\phi}}\right)^{\frac{1}{4}} \rightarrow t^{\frac{1}{2}}$. Thus, setting ${\rm SNR}=1$ in Eq.~(\ref{SNRvolt}) with a 30 days of integration time, we find the sensitivity of this experiment, see Fig.~\ref{FigCapSens}.

As compared with other experiments \cite{delay,GEO600,Aharony,Oswald2021,Tretiak}, detection of the scalar field dark matter with the capacitor may have advantages for various dark matter masses. In particular, the projected sensitivity is nearly two orders in magnitude higher than that of the molecular spectroscopy experiment \cite{Oswald2021} and is comparable to the atomic spectroscopy experiment \cite{Tretiak} and unequal-delay interferometer experiment \cite{delay}. Although the capacitor experiment is not as sensitive as the gravitational wave detectors \cite{GEO600}, it would allow one to probe the region of higher scalar field mass. It should also be noted that limits from all these experiments are significantly weaker than the constraints from the equivalence principle test \cite{EP1,EP2,Stadnik-equivalence}. Nevertheless, they provide important information about this dark matter model because they impose independent constraints on the couplings.

Note that the capacitor experiment proposed in this section may be used for probing the axion electromagnetodynamics model \cite{Sokolov}. In particular, all limits on $g_{\phi\gamma\gamma}$ should apply to the coupling constant $g_{a{\rm AB}}$ in this model as well.


\subsection{Limits on $g_{\phi\gamma\gamma}$ from ADMX}
\label{SecADMX}

\begin{figure}
    \centering
    \includegraphics[width=8.5cm]{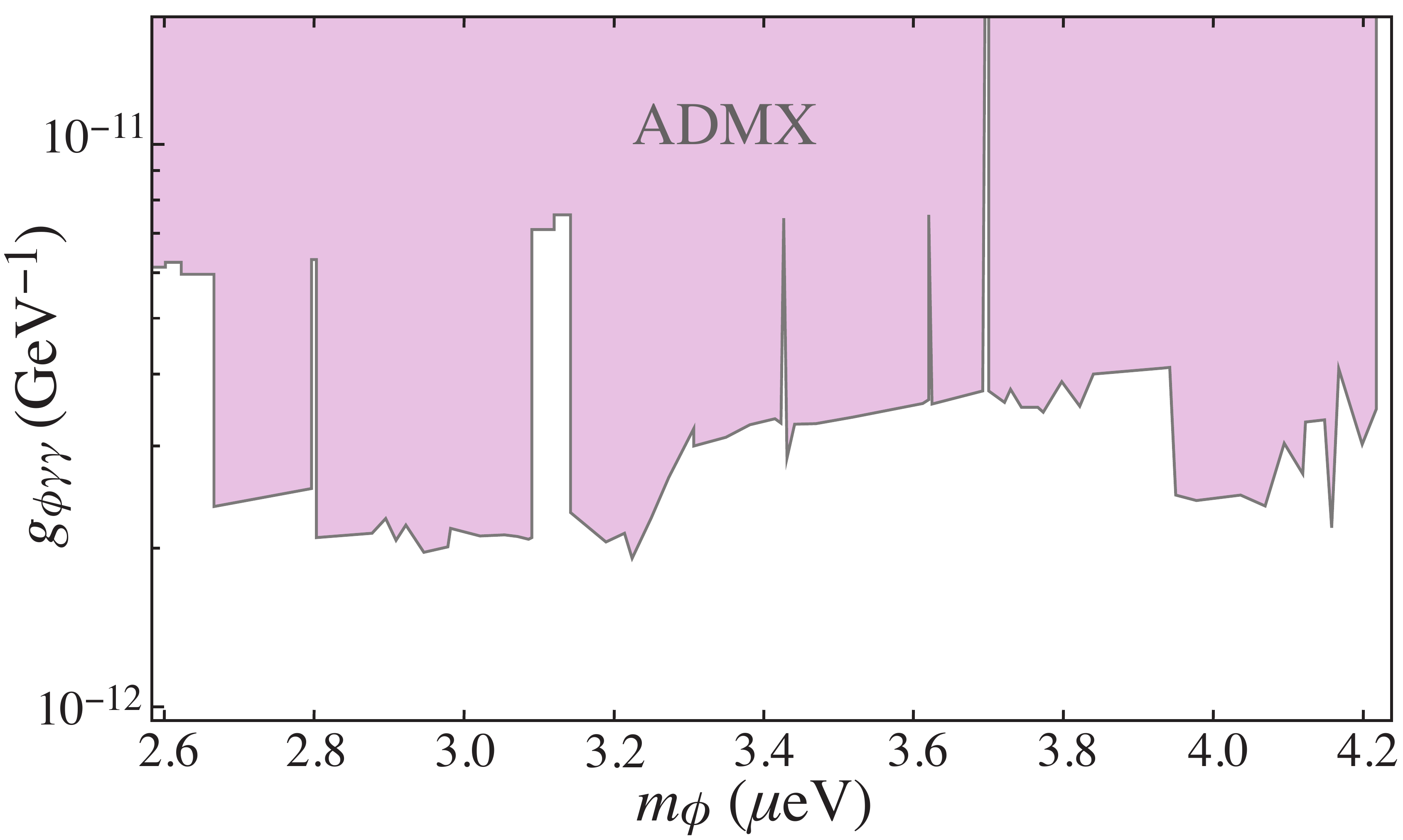}
    \caption{Upper limits on the coupling $g_{\phi\gamma\gamma}$ as a function of mass $m_\phi$. These limits are found with a graphical accuracy by re-scaling with the constant (\ref{kappa}) the corresponding constraints on the axion coupling constant $g_{a\gamma\gamma}$ reported in Refs.~\cite{ADMX6,ADMX9}.}
    \label{FigExclusion}
\end{figure}

In typical haloscope experiments, a TM$_{010}$ mode in a cylindrical cavity is employed, inside a solenoidal magnetic field aligned along the $z$-axis of the cavity. If the applied magnetic field is perfectly uniform along the $z$-direction with no fringing, then the scalar dark matter form factor~\eqref{CB} is exactly zero for such modes in cylindrical cavities. In fact, it does not matter how one orients the cavity with respect to the applied magnetic field, the form factors for all modes inside a typical (haloscope-style) cylindrical cavity, with a perfectly uniform axial magnetic field are zero.

In reality, solenoid fields are not perfectly uniform, and fringing effects exist, which create small but non-zero form factors for various modes inside  resonators. For instance, for the TM$_{010}$ mode in the ADMX Run 1c experiment, with the real ADMX solenoid, the form factor \eqref{CB} is estimated as
\begin{equation}
\label{C40}
    C_\alpha \approx 10^{-12}\,,
\end{equation}
and for the ADMX Run 1c Sidecar experiment \cite{ADMXsidecar}
\begin{equation}
\label{C-Run-1c}
    C_\alpha \approx 10^{-8}\,.
\end{equation}
This value imposes a strong suppression as compared with the axion detection form factor $C_{\rm axion} = 0.4$ in the ADMX experiment \cite{ADMX6,ADMX9}. Nevertheless, it is still possible to derive new constraints on the scalar-photon coupling $g_{\phi\gamma\gamma}$ by re-scaling the constraints on the axion-photon coupling reported in Refs.~\cite{ADMX6,ADMX9} by the factor 
\begin{equation}
\label{kappa}
\kappa \equiv \sqrt{C_\alpha/0.4} = 1.6\times 10^{-4}\,.
\end{equation}
Here we assumed the value (\ref{C-Run-1c}) for the scalar field detection form factor. The corresponding limits on $g_{\phi\gamma\gamma}$ in Fig.~\ref{FigExclusion} vary from 
\begin{equation}
   g_{\phi\gamma\gamma} \lesssim 2.5\times 10^{-12}\ {\rm GeV}^{-1}\    \mbox{ for }   m_\phi = 2.7\mu{\rm eV}
\end{equation}
to
\begin{equation}
    g_{\phi\gamma\gamma} \lesssim 3.5\times 10^{-12}\ {\rm GeV}^{-1}\
    \mbox{ for } m_\phi= 4.2\,\mu{\rm eV}.
\end{equation}
To the best of our knowledge, the coupling constant $g_{\phi\gamma\gamma}$ has not been constrained in this region yet. We hope that future experiments may further improve these constraints.

Note that in the calculation of the form factors (\ref{C40}) and (\ref{C-Run-1c}) we ignored the factor $e^{i\vec p\cdot\vec x}$ in Eq.~(\ref{CB}) because $\vec p\cdot\vec x \ll1$. However, the terms with $\vec p\cdot\vec x$ introduce the dependence of the from factor on the scalar field momentum and may be responsible for daily and annual modulations of the signal. Moreover, such terms may produce non-vanishing form factors for some of the modes in the cavity. In particular, we find that the TE$_{111}$ mode has a form factor $C\approx 1.7\times 10^{-6}$ when the momentum $p$ is aligned in the $x$ direction of the cavity. Given this value of the form factor, the constraints in Fig.~\ref{FigExclusion} could be further improved if this mode would be measured in the ADMX experiment.

\subsection{New cavity resonator proposals}
\label{SecNewCavities}

\begin{figure}
    \centering
    \includegraphics[width=8.5cm]{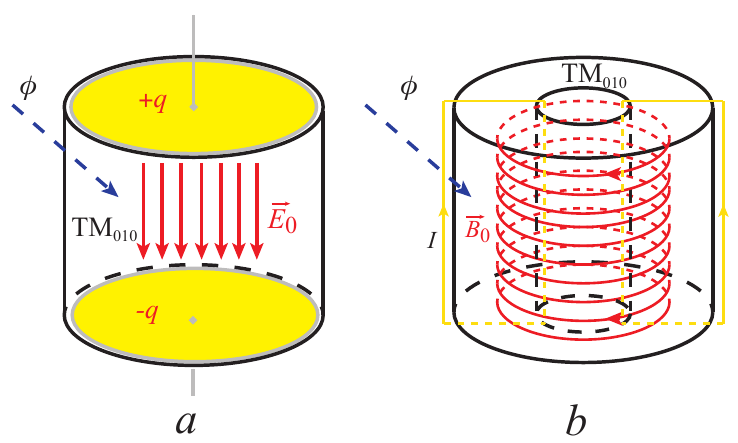}
    \caption{Designs of cavity resonator experiments for detection of the scalar field dark matter. (a) A cavity utilizing a strong permanent electric field $\vec E_0$. (b) A ring-shaped cylindrical cavity with azimuthal magnetic field $\vec B_0$ (red circles) created by a toroidal solenoid with current $I$ (yellows loops).}
    \label{fig1}
\end{figure}

Equations (\ref{C40}) and (\ref{C-Run-1c}) show that the cavity resonator in the ADMX experiment has low sensitivity to scalar field coupling constant $g_{\phi\gamma\gamma}$. In this section, we discuss some proposals of cavity resonators with maximized sensitivity to the scalar field dark matter.  

Equations (\ref{P33}) and (\ref{formfactor}) show that the role of electric and magnetic fields is swapped as compared with the corresponding expressions in the axion cavity experiment proposals \cite{Sikivie83,Sikivie85}. Therefore, to maximize the sensitivity to the scalar field dark matter one may consider a cylindrical cavity resonator with an axial static electric field $\vec E_0$, see Fig.~\ref{fig1}(a). In such a resonator, top and bottom caps play role of the capacitor plates. Assuming that the strength of the applied electric field may be of order $E_0\sim 1$ MV/m, the estimate of the signal power is given by Eq.~(\ref{PowerElectric}).  The form factor (\ref{P38-formfactor}) is maximized for the TM$_{010}$ mode, $C_\alpha=0.69$ for a perfect cylindrical cavity.

Cavity resonators permeated by the magnetic field are commonly used in the searches of wavy dark matter because it is technically challenging to create strong electric fields in a large volume. In Fig.~\ref{fig1}(b), we propose a magnetic-field-based cavity resonator with maximized sensitivity to the scalar field dark matter. Such a cylindrical cavity should have a shape of a ring permeated by an azimuthal magnetic field, which could be applied with a toroidal solenoid, where current loops run through the central hole in the cavity, and around the outside. In such a cavity, TM$_{010}$ modes are present, but unlike in the axial magnetic field they have appreciable scalar dark matter form factors, given the applied magnetic field runs in the $\varphi$ direction (the same direction as the cavity mode magnetic field). Numerical simulations with the use of COMSOL software indicate that such cavities can achieve form factors close to unity, $C_\alpha\sim1$, but exact value of this form factor depends on the cavity parameters. This makes the cavity resonator in Fig.~\ref{fig1}(b) very promising for future experiments searching for the scalar field dark matter with sub-meV mass.


\section{Summary and discussion} 
\label{SecSummary}

In this paper, we developed a theory and a number of new experimental proposals for measuring the constant of scalar-photon interaction $g_{\phi\gamma\gamma}$. This coupling constant quantifies the strength of the dilaton-like interaction of the scalar field dark matter with the electromagnetic field (\ref{L1}) that is analogous to the axion-photon coupling $g_{a\gamma\gamma}$. In similarity with the axion case \cite{Sikivie83,Sikivie85}, the scalar field dark matter can transform into photons in strong electric and magnetic fields. We found the power of the corresponding photon signal (\ref{Power}) in the case of non-resonant process and (\ref{P33}) in cavity resonators. In comparison with the axion case \cite{Sikivie83,Sikivie85} (see also \cite{Sikivie2020}), the roles of electric and magnetic fields in these expressions are swapped. As a consequence, existing cavity-based axion dark matter search experiments like ADMX or ORGAN have a low sensitivity to the scalar-photon coupling constant $g_{\phi\gamma\gamma}$. The experiments searching for Solar axions like CAST, however, possess a good sensitivity to scalar-photon coupling $g_{\phi\gamma\gamma}$. This allows us to find new limits on this coupling by re-purposing the results of the CAST experiment, see Fig.~\ref{FigCAST}.

We demonstrate that a perfect cylindrical cavity permeated by uniform axial magnetic field does not allow one to detect the scalar field dark matter because the corresponding cavity form factor (\ref{CB}) vanishes identically. In actual cavities, however, the solenoid magnetic field is not perfectly uniform, and the form factor (\ref{CB}) acquires a small but non-vanishing value. In particular, due to the fringing of the magnetic field in the ADMX experiment \cite{ADMX6,ADMX9} the form factor is of order $C_\alpha \approx 10^{-12}$ while in the ADMX Sidecar experiment \cite{ADMXsidecar} it is $C_\alpha \approx 10^{-8}$, see Eqs.~(\ref{C40}) and (\ref{C-Run-1c}), respectively. These values of the form factor allow us to find new limits on the coupling $g_{\phi\gamma\gamma}$ in the scalar field mass range from 2.6 to 4.2 $\mu$eV, see Fig.~\ref{FigExclusion}. These limits are obtained from the corresponding constraints on the axion-photon coupling $g_{a\gamma\gamma}$ by re-scaling the results of the ADMX experiment \cite{ADMX6,ADMX9} with the coefficient (\ref{kappa}).

To enhance the value of the form factor for scalar field dark matter detection some modifications of the existing cavity resonators are needed. In Sec.~\ref{SecNewCavities} we propose a new ring-shaped cavity resonator permeated with azimuthally oriented magnetic field induced by a toroidal solenoid. As we show, the from factor of such a cavity appears of order $O(1)$, although exact value depends on particular geometry of the cavity. Alternatively, a high value of the form factor may be achieved in a cylindrical cavity permeated by a strong axial electric field, $C_\alpha=0.69$. We hope that either of these proposals may be implemented in future experiments searching for scalar field dark matter.

Although cavity resonators may have a good sensitivity to light scalar field dark matter, this detection technique does not allow one to cover a wide range of masses of this particle. In Sec.~\ref{SecBroadband} we propose a broadband detection technique utilizing a capacitor charged to a high voltage. The scalar field interacting with a static electric field of the capacitor induces an effective polarization and electric field in the capacitor which oscillate with the scalar field frequency. The signal in this experiment is formed by the corresponding oscillating component of the voltage on the plates of the capacitor. Assuming reasonable values for the external voltage source and the noise level in the detector, we show that the sensitivity to $g_{\phi\gamma\gamma}$ in this experiment may be nearly two orders in magnitude higher than the existing constraints on this coupling from the molecular spectroscopy experiment \cite{Oswald2021} and is comparable to the atomic spectroscopy experiment \cite{Tretiak}, see Fig.~\ref{FigCapSens}. The advantage of this experiment is that the signal (\ref{signal}) has linear dependence on the coupling constant $g_{\phi\gamma\gamma}$. This makes this technique very promising for searches of the light scalar field dark matter. 

\vspace{3mm}
\textit{Acknowledgements} --- 
We thank Yevgeny Stadnik for useful references and comments. IBS is very grateful to Dmitry Budker, Gilad Perez and Oleg Tretiak for clarifying discussions on the details of atomic and molecular spectroscopy experiments and constraints from the equivalence principle. The work of VVF and IBS was supported by the Australian Research Council Grants No. DP190100974 and DP200100150 and the Gutenberg Fellowship. The work of MET and BTM was supported by the Australian Research Council Centre of Excellence for Engineered Quantum Systems, CE170100009 and Centre of Excellence for Dark Matter Particle Physics, CE200100008. BTM is also supported by the Forrest Research Foundation. The work of IBS was supported in part by the Alexander von Humboldt foundation. CAD Models for the ADMX cavity and magnetic field map were supplied by the ADMX Collaboration.


%

\end{document}